\documentclass[aps,showpacs,twocolumn]{revtex4}
\usepackage{amsmath, amsthm,times}
\usepackage{epsfig}
\usepackage{amssymb}
\usepackage{graphicx}

\begin{document}

\sloppy

\title{Characterizing Entanglement Sources}

\author{Pavel Lougovski$^{1}$ and S.J. van Enk$^{1,2}$}

\address{
$^1$Department of Physics and Oregon Center for Optics, University of Oregon\\
 Eugene, OR 97403\\
$^2$Institute for Quantum Information, California Institute of Technology, Pasadena, CA 91125}
\begin{abstract}
We discuss how to characterize entanglement sources with finite sets of measurements. The measurements do not have to be tomographically complete, and may consist of POVMs rather than von Neumann measurements. Our method yields a probability that the source generates an entangled state as well as estimates of any desired calculable entanglement measures, including their error bars.  We apply two criteria, namely Akaike's information criterion  and the Bayesian information criterion, to compare and assess different models (with different numbers of parameters) describing  entanglement-generating devices. We discuss differences between standard entanglement-verificaton methods and our present method of characterizing an entanglement source.
\end{abstract}


\maketitle

\section{Introduction}

Entanglement is useful, but hard to generate, and even harder to detect. The most measurement-intense approach to the problem of experimentally detecting the presence of entanglement is to perform complete quantum-state tomography \cite{tomo}. Even for just two qubits this implies a reconstruction of all 15 independent elements of the corresponding density matrix. Subsequently applying the positive partial transpose (PPT) criterion  to the reconstructed matrix gives a conclusive answer about entanglement or separability of the state~\cite{Witness,Peres}.

From the practical point of view it is desirable to have an entanglement detection tool that is more economical than full state tomography but nevertheless is decisive. Already in the original work on PPT~\cite{Witness} it was noticed that one can always construct an observable $\mathcal{W}$ with non-negative expectation values for all separable states $\rho_{s}$ and a negative expectation value for at least one entangled state $\rho_e$. In this way an experimentally detected violation of the inequality $\langle\mathcal{W}\rangle \ge 0$ is a sufficient condition for entanglement. The observable $\mathcal{W}$ is called an entanglement witness (EW). There always exists an optimal choice of {\em local} orthogonal observables such that a given EW can be expressed as a sum of their direct products~\cite{Guehne}, so that a witness can always be measured locally. The advantage of using EWs for entanglement detection will be appreciated better for multi-partite systems with  more than two qubits, because the number of tomographic measurements would grow exponentially with the number of qubits. On the other hand, a given witness does not detect all entangled states and therefore a variety of different EWs should be tested in order to rule out false negative results.

EWs assume the validity of quantum mechanics, and also assume one knows what measurements one is actually performing. A valuable alternative to EW can be sought in using a violation of Bell-CHSH inequalities \cite{bell,chsh} as a sufficient condition for entanglement (although a Bell-inequality test can be formulated as a witness, too \cite{Bellwitness}). Because Bell inequalities are derived from classical probability theory without any reference to quantum mechanics, no assumption about what is being measured is necessary. This method is, therefore, safe in the sense of avoiding many pitfalls arising from unwarranted (hidden) assumptions about one's experiment \cite{entmeasurement}.

Here we propose a different method for characterizing an entanglement source that automatically takes into account finite data as well as imperfect measurements.
Our method consist of two parts. The first part, ``Bayesian updating,'' produces an estimate of the relative probabilities that entangled and separable states are consistent with a given finite set of data. This estimate depends on what {\em a priori} probability distribution (the {\em prior}) one chooses over all possible states (the more data one has, the less it depends on the prior). That is, there is an {\em a priori} probability of entanglement, and each single measurement updates this probability to an {\em a posteriori} probability of entanglement. The latter then has to be compared to the former, in order to reach the conclusion that one is now either more certain or less certain about having produced an entangled state. In fact, every experiment can only make such probabilistic statements about entanglement, although this is almost never explicitly stated in these terms.
Thus our method differs from those in
Refs~\cite{AudenaertWitness,JensWitness,GuehneWitness} which assume expectation values of EWs are known [corresponding effectively to an {\em infinite} data set] and try to find the {\em minimally}-entangled state consistent with those expectation values.
We use a numerical Bayesian updating method for a probability distribution over density matrices, which is similar to that recently discussed in Ref.~\cite{Blume-Kohout} in the context of quantum-state tomography.  In particular, whereas
the reconstruction of a density matrix from experimental data is usually based on the maximum likelihood estimation (MLE),  Ref.\cite{Blume-Kohout} discusses its drawbacks and proposes Bayesian updating in its stead as a superior method.  Our aim, though, is not to give an estimate of the density matrix, but of entanglement. In fact, any quantity that can be calculated from a density matrix, such as the purity of one's state, can be estimated this way.

The second part of our method introduces two information criteria \cite{book} to judge how different models of a given entanglement generation process can be compared to each other quantitatively. It is probably best to
explain this part by giving an example. For simplicity, we consider the case of two-qubit states.
Suppose an experimentalist has a model for her entanglement generating source that contains, say, two parameters describing two physically different  sources of noise in the final two-qubit state produced.
She may try to fit her data to her two-parameter model, but obviously  there are always states in the full 15-dimensional set of all physical states that will fit the data better. There are a number of criteria, standard in the literature on statistical models, that compare quantitatively how different models fit the data. Here we will use Akaike's Information Criterion (AIC) and the Bayesian Information Criterion (BIC) \cite{book}. These information criteria aim to find the most informative model, not the best-fitting model.
The idea is that a two-parameter model fitting the data almost as well as the full quantum-mechanical description  would provide more physical insight and a more economical (think Occam's razor) and transparent description.  Each of the two information criteria, AIC and BIC,
produces a number $\Omega$.
One term in $\Omega$ is the logarithm of the maximum likelihood possible within each model, and the second term subtracts a penalty for each parameter used in the model. The model with the larger value of $\Omega$ is then deemed to be the more informative.
We propose here to combine information criteria with the Bayesian updating methods for entanglement estimation. Namely, we propose to use the more informative model to generate a ``substitute prior.'' In the case that the simpler model is the more informative, the numerical efforts required for our Bayesian updating method are much smaller, and yet should lead to correct descriptions of the entanglement generated by one's source.

This paper is organized as follows. In Section \ref{Bayesup}
we give a general formulation of our method of Bayesian updating applicable to any quantum system.  We also  formulate precisely the two information criteria for model selection. In Section \ref{exmpl}
we discuss numerical examples, which  illustrate the Bayesian methods and the information criteria. For concreteness we consider measurements of Bell-CHSH correlations (although any sort of measurements would do).
The examples show that our method detects entangled two-qubit states that escape detection by any of the Bell-CHSH inequalities and even by violations of the stronger version of these bounds, which we call Roy-Uffink-Seevinck bounds \cite{roy, UffinkSeevinck,quantph}. In the Discussion and Conclusions Section we discuss the essential difference between our method of characterizing an entanglement source and the standard methods of entanglement verification \cite{guehne,entmeasurement}.

\section{Quantifying Entanglement via Bayesian Updating}\label{Bayesup}

Here we present  a numerical Bayesian updating method for one's probability distribution over density matrices. A related Bayesian method was recently advocated in Ref.~\cite{Blume-Kohout} in the context of quantum-state tomography and quantum-state estimation.  We note our aim is not to give an estimate of the density matrix, but, more modestly, to give estimates of entanglement, purity, and in principle any quantity that can be efficiently calculated from the density matrix. We first discuss the method in general, and subsequently we propose a new method to choose a prior probability distribution over density matrices.
\subsection{Method}
The method itself can be formulated as a five-step procedure:
\begin{enumerate}
\item For a system of $M$ qubits we first choose a  finite test set of density matrices. We calculate the amount of entanglement (in fact, the negativity) for each state in the set \footnote{We remark that in higher dimensions (or for more than two parties) the negativity does not necessarily pinpoint all classes of entangled states. In this case more than one entanglement monotone should be used to characterize entanglement.}. The {\em a priori} probability that our unknown experimentally generated state, which we denote by $\rho_{?}$, equals  a state $\rho$ in the set is chosen as ${\rm p}_{{\rm prior}}(\rho)=1/N_{{\rm s}}$, where $N_{{\rm s}}$ is the number of states in the set.

\item We assume some set of POVMs with elements $\{\Pi_i\}$ is measured. These POVMs can describe any (noisy) set of measurements one performs on the qubits.

\item For the acquired measurement record $d = \{d_{1},\cdots,d_{i}\}$ consisting of the number of times outcome $i$ was obtained \footnote{We thus implicitly assume identical and independent copies of states of $M$ qubits.}, we calculate the quantum-mechanical probability ${\rm p} (d|\rho)$ that a given state $\rho$ from the test set generates the measurements outcome $d$ (which follows directly from Tr$\rho\Pi_i$).  Having at hand probabilities ${\rm p} (d|\rho)$ for all states $\rho$ in the test set we are now able to calculate the {\em a priori} probability ${\rm p}(d)=\sum\limits_{\rho}{\rm p} (d|\rho)/N_{{\rm s}}$ for the measurement record $d$ to occur.

\item We calculate -- using Bayes' rule -- the probability ${\rm p}(\rho|d)$ of having the state $\rho$ given the measurement outcomes $d$: ${\rm p} (\rho|d) ={\rm p} (d|\rho)/[N_{{\rm s}}{\rm p} (d)]
$.

\item We obtain the posterior probability distribution over density matrices in our test set: ${\rm p}_{{\rm posterior}} (\rho):={\rm p} (\rho|d)$ for all states $\rho$.
\end{enumerate}
We can then repeat steps 2-5 for a new set of measurements $d$, if needed.

This procedure gives us, in step 5, a numerical estimate of the {\em a posteriori} probability  that the unknown state $\rho_{?}$ equals the state $\rho$ from the test set. From p$(\rho|d)$ we can estimate the probability ${\rm p}_{e}$ for the state $\rho_{?}$ to be entangled. We just sum the probabilities ${\rm p} (\rho|d)$ for all entangled states $\rho_{ent}$ in the set i.e.
\begin{equation}
{\rm p_e} (\rho_{?}) = \sum\limits_{\rho=\rho_{ent}} {\rm p} (\rho|d).
\end{equation}
Furthermore, we can calculate probability distributions for any function of the density matrix, such as the negativity and purity. We thus infer expectation values such as
\begin{equation}
\overline{N} = \sum\limits_{\rho=\rho_{ent}} {\rm p} (\rho|d)
N(\rho),\end{equation}
 and
  \begin{equation}
\overline{P } = \sum\limits_{\rho} {\rm p} (\rho|d)
{\rm Tr}(\rho^2),
\end{equation}
as well as standard deviations $\sigma_N=\sqrt{\overline{N^2}-\bar{N}^2}$ etc.

The meaning of our final probability distribution p$(\rho|d)$ and of the above expectation values is as follows. If we were forced to give a {\em single} density matrix that best describes  all data and that includes error bars,  we would give the mixed state $\bar{\rho} = \int d\rho {\rm p}(\rho|d)(\rho)\rho$, as explained in \cite{Blume-Kohout}. The purity and negativity of the state $\bar{\rho}$ are {\em not} equal to (in fact, smaller than) the estimates $\bar{N}$ and $\bar{P}$ that we use here. The difference is this: if one were to perform more measurements that are tomographically complete, $\bar{N}$ is the expected negativity of the final estimated density matrix. $N(\bar{\rho})$, on the other hand, would be the useful entanglement of a single copy available {\em without} performing more measurements. For most quantum information processing purposes (such as teleportation) one indeed needs more precise knowledge about the density matrix than just its entanglement.  See Ref.~\cite{entmeasurement} for more discussions on this issue.
\subsection{Model testing and information criteria}
The only problem standing in the way of a straightforward application of the above Bayesian updating procedure is that a sufficiently dense test set (used in step 1) is in general too hard to handle numerically, since even for  two-qubit density matrices the parameter space is 15-dimensional.
Although there are certainly ways out of this problem (in particular, sampling directly from the posterior probability distribution can be efficiently done with the Metropolis-Hastings algorithm, see e.g. \cite{MH}), here we stick to the idea of a set of test states by simplifying that set, as follows.

As an illustrative example (which we will again consider in great detail in the next Section), consider an experimentalist trying to produce a maximally entangled Bell state of 2 qubits, say, $(|00\rangle+|11\rangle)/\sqrt{2}$. She wants to test her entanglement-generating device by measuring some set of Bell correlations. In particular, suppose she measures $2<K<15$ independent observables.

From her previous experience with the same device, she models the generation process by assuming there is both Gaussian phase noise and white noise (mixing with the maximally mixed state $\propto \openone$). That is, she assumes her device generates  states of the form
\begin{equation}\label{ps1}
\rho_{p,\sigma}=p\rho_\sigma+(1-p)\frac{\openone}{4},
\end{equation}
with $p\in [0,1]$ and
\begin{equation}\label{ps2}
\rho_\sigma=\frac{1}{2}\int_{-\pi}^{\pi} d\phi P(\phi)
(|00\rangle+\exp(i\phi)|11\rangle)
(\langle 00|+\exp(-i\phi)\langle 11|).
\end{equation}
Here $P(\phi)$ is a Gaussian phase distribution of the form
\begin{equation}\label{ps3}
P(\phi)=N_\sigma\exp(-\phi^2/\sigma^2)
\end{equation}
with the normalization factor $N_\sigma$ given by
\begin{equation}
N_\sigma=\frac{1}{\int_{-\pi}^\pi d\phi
\exp(-\phi^2/\sigma^2)}.
\end{equation}
So there are just two parameters the experimentalist has to determine from her measurement results, $p$ and $\sigma$.

As a measure to judge how well her data $d$ fit the model (\ref{ps1})--(\ref{ps3}), she considers the best likelihood for that model,
\begin{equation}
L_{p,\sigma}\equiv\max_{p,\sigma}P(d|\rho_{p,\sigma}).
\end{equation}
She would like to compare this to the maximum likelihood over {\em all} physical two-qubit states $\rho$,
\begin{equation}
L_a\equiv \max_{\rho}P(d|\rho).
\end{equation}
There are several ways to compare these two quantities \cite{book}. One criterion is called
Akaike's Information Criterion, and it defines the quantity
\begin{equation}
\Omega=\log(L)-k
\end{equation}
for each model, where $k$ is the number of parameters in the model, and $L$ is the maximum likelihood for the model. The quantity $\Omega$ rewards a high value of the best likelihood (indicating a good fit), but penalizes a large number of parameters (to guard against overfitting).
Now when measuring $2<K<15$ observables, the best complete model contains just $K$, not 15, independent parameters.

Thus the experimentalist would calculate two numbers
\begin{eqnarray}
\Omega_{p,\sigma}&=&\log (L_{p,\sigma})-2,\nonumber\\
\Omega_{a}&=&\log (L_a)-K.\label{Omega}
\end{eqnarray}
If $\Omega_{p,\sigma}>\Omega_a$ then
the Akaike Information Criterion judges the simple 2-parameter model to be more informative than the complete $K$-parameter model.

There is a Bayesian version of this criterion \cite{book}, and it is defined in terms of similar quantities
\begin{equation}
\Omega'=\log(L)-k\log(N_{{\rm m}})/2,
\end{equation}
where $L$ and $k$ have the same meaning as before, and $N_{{\rm m}}$ is the number of data taken.
Again, if $\Omega'_{p,\sigma}>\Omega'_m$, the 2-parameter model is considered more informative than the $K$-parameter description. For $N_{{\rm s}}>8$ the BIC puts a larger penalty on the number of parameters than does the AIC.

In the case that the simple model turns out to be more informative, according to at least one of the two criteria [this depends on the data], we propose that the experimentalist may well use the simple model to construct a test set of states.
For example, she could assume as prior probability distributions for $p$ and $\sigma$ that $p$ is uniformly distributed on the interval $[0,1]$, and that $\sigma$ is uniform on, say, the interval $[0,\pi]$ (this is somewhat arbitrary, of course, as every prior is). Then, the test set of
states could be sampled by simply choosing
$N_p$ uniformly spaced points in the
interval $[0,1]$ for $p$ and $N_\sigma$ uniformly spaced points in the interval $[0,\pi]$ for $\sigma$, thus creating a test set of $N_{{\rm s}}=N_p\cdot N_\sigma$ states.

The above model leads to states that are diagonal in the Bell basis,
\begin{eqnarray}
|\Phi_1\rangle&=&(|00\rangle+|11\rangle)/\sqrt{2}\nonumber\\
|\Phi_2\rangle&=&(|00\rangle-|11\rangle)/\sqrt{2}
\nonumber\\
|\Phi_3\rangle&=&(|01\rangle+|10\rangle)/\sqrt{2}
\nonumber\\
|\Phi_4\rangle&=&(|01\rangle-|10\rangle)/\sqrt{2}
\end{eqnarray}\label{Bb}
In the next Section we thus consider not only
the above two-parameter model, but also its obvious extension to a three-parameter model by allowing all Bell-diagonal states.

\section{Examples}\label{exmpl}

\subsection{Orthogonal spin measurements}
In the following we consider, as an example, two-qubit states with spin measurements performed on each qubit (considered as a spin-1/2 system)  in {\em two} arbitrary  spatial directions that are orthogonal, and we denote the corresponding spin operators by $A_1$ and $A_2$ for the first qubit, and $B_1$ and $B_2$ for the second qubit. This consitutes a measurement of 8 independent quantities,
four single-qubit expectation values and four correlations.

We note that in this case we can construct four Bell-CHSH operators from the four measured correlations:
\begin{eqnarray}
\mathcal{B}_1&:=& A_{1}\otimes(B_{1} + B_{2}) +A_{2}\otimes(B_{1} - B_{2}), \nonumber\\
\mathcal{B}_2&=& A_{1}\otimes(B_{1} + B_{2}) -A_{2}\otimes(B_{1} - B_{2}), \nonumber\\
\mathcal{B}_3&=& A_{1}\otimes(B_{1} - B_{2}) + A_{2}\otimes(B_{1} + B_{2}), \nonumber\\
\mathcal{B}_4&=& A_{1}\otimes(-B_{1} + B_{2}) + A_{2}\otimes(B_{1} + B_{2}).\label{AllBell}
\end{eqnarray}
We then test two-qubit states that may be entangled but that do {\em not} violate any of the four Bell inequalities that can be constructed from these four operators.
In fact, we will not even optimize the choice of spatial directions, given an initial guess of what state should be produced, for violating a Bell inequality.

Finally, we will add one more correlation to be measured, namely that involving the third dimension: $A_3 B_3$. That is, whenever $A_3$ is measured on the first qubit, $B_3$ is measured on the second qubit. This addition makes the measurements on each qubit separately tomographically complete, but it does not lead to additional Bell-CHSH operators. The total number of independent observables measured in this case is 11 (four are missing).
Thus, the parameter $K$ to be used for evaluating $\Omega_a$ of Eq.~(\ref{Omega}) is  $K=11$.

\subsection{Analytical results}
Determining the AIC and BIC criteria can be done analytically in most cases that we will consider here.
First of all, we can bound the maximum likelihood over all states, given the sort of measurements from the preceding subsection.
There are 20 observed frequencies, as follows:
for each of the five correlation measurement $A_i B_j$, where
$i,j$ take on the values $(i,j)=(1,1), (1,2), (2,1), (2,2), (3,3)$
there are four different outcomes, which we can denote
by $(+,+), (+,-), (-,+), (-,-)$. If we denote these frequencies by $f_{ijk}$, for $k=1\ldots 4$, then the (log of the) maximum likelihood is bounded by
\begin{equation}
\log (L_a)\leq
\sum_{k,(ij)} N_{ij}f_{ijk}\log(f_{ijk}),
\end{equation}
where $N_{ij}$ is the number of times
the $A_iB_j$ correlation was measured.
If we assume all five correlations are measured equally often, then we have
\begin{equation}
\log (L_a)\leq\frac{ N_{{\rm m}}}{5}
\sum_{k,(ij)}f_{ijk}\log(f_{ijk}),
\end{equation}
The bound is achieved when
there is a physical state predicting the  frequencies exactly as they were observed.

Let us choose directions of our spin measurements as
$A_1=B_1=X$, $A_2=B_2=Y$ and $A_3=B_3=Z$.
Then, there are two obvious models an experimentalist could choose from: the first is the Bell-diagonal model, containing three parameters, in which states are of the form
\begin{equation}
\rho=\sum_{i=1}^4 p_{i} |\Phi_{i}\rangle\langle \Phi_{i}|.
\end{equation}
The observed frequencies $f_{ijk}$ for the five correlations
cannot be all predicted  to arbitrary accuracy by Bell-diagonal states. In fact,
most frequencies predicted by this model are independent of the values of $\{p_i\}$, and are equal to 1/4.
The only predicted frequencies (which we denote by $\tilde{f}$ so as to distinguish them from the observed frequencies $f$) that actually depend on the values of $\{p_i\}$ are
\begin{eqnarray}
\tilde{f}_{111}&=&\tilde{f}_{114}=p_1/2+p_3/2,\nonumber\\
\tilde{f}_{221}&=&\tilde{f}_{224}=p_2/2+p_3/2,\nonumber\\
\tilde{f}_{331}&=&\tilde{f}_{334}=p_1/2+p_2/2,\nonumber\\
\tilde{f}_{112}&=&\tilde{f}_{113}=p_2/2+p_4/2,\nonumber\\
\tilde{f}_{222}&=&\tilde{f}_{223}=p_1/2+p_4/2,\nonumber\\
\tilde{f}_{332}&=&\tilde{f}_{333}=p_3/2+p_4/2.
\end{eqnarray}
The best-fitting Bell-diagonal state can only predict
the correct correlations between $XX$, $YY$, and $ZZ$ measurements.
For example, there is a Bell-diagonal state predicting the correct value for the sum $\tilde{f}_{111}+\tilde{f}_{114}$, but its prediction for the difference will always be zero. Thus, the Bell-diagonal state fitting the data best will have the following values for $\{p_i\}$:
\begin{eqnarray}
p_1&=&[f_{111}+f_{114}-f_{221}-f_{224}+f_{331}+f_{332}]/2,\nonumber\\
p_2&=&[-f_{111}-f_{114}+f_{221}+f_{224}+f_{331}+f_{332}]/2,\nonumber\\
p_3&=&[f_{111}+f_{114}+f_{221}+f_{224}-f_{331}-f_{332}]/2,
\end{eqnarray}
provided the observed frequencies are such that the $\{p_i\}$ including $p_4=1-p_1-p_2-p_3$ are all nonnegative. In that case, the (log of the) maximum likelihood over all Bell-diagonal states is
 \begin{eqnarray}
\log L_{Bd}&=&\frac{ N_{{\rm m}}}{5}\left[\sum_i
(f_{ii1}+f_{ii4})\log([f_{ii1}+f_{ii4}]/2)\right.\nonumber\\
&&+\sum_i(f_{ii2}+f_{ii3})\log([f_{ii2}+f_{ii3}]/2)\nonumber\\
&&+\left.\sum_{k,i\neq j}f_{ijk}\log(1/4)\right]
\end{eqnarray}
For the Bell-diagonal model
we can construct a prior distribution over Bell-diagonal states by choosing the numbers $\{p_i\}$
uniformly over the simplex, as explained in \cite{NegativityOriginal}.

The two-parameter model is similar in its predictions the the Bell-diagonal model. The only difference is that
the parameters $p_3$ and $p_4$ are equal.
Thus this model can predict only two correlations correctly, namely $ZZ$ and $XX-YY$.
The maximum likelihood for this model, then, is given by
 \begin{eqnarray}
\log L_{p,\sigma}&=&\frac{ N_{{\rm m}}}{5}[
(f_{331}+f_{334})\log([f_{331}+f_{334}]/2)\nonumber\\
&&+(f_{332}+f_{333})\log([f_{332}+f_{333}]/2)\nonumber\\
&&+(f_{111}+f_{114}+f_{222}+f_{223})\times\nonumber\\
&&\log(1/4+(f_{111}+f_{114}-f_{221}-f_{224})/2)\nonumber\\
&&+(f_{221}+f_{224}+f_{112}+f_{113})\times\nonumber\\
&&\log(1/4+(f_{221}+f_{224}-f_{111}-f_{114})/2)\nonumber\\
&&+\sum_{k,i\neq j}f_{ijk}\log(1/4)],
\end{eqnarray}
provided all inferred frequencies are nonnegative.

The three parameters to be used for selecting the most informative model are then, in the case of AIC:
\begin{eqnarray}
\Omega_a&=&\log L_a-11,\nonumber\\
\Omega_{p,\sigma}&=&\log L_{p,\sigma}-2\nonumber\\
\Omega_{Bd}&=&\log L_{Bd}-3,
\end{eqnarray}
and similar expressions for the BIC.

\subsection{Numerics}
\begin{figure}[htbp]
\begin{center}
\includegraphics[width=3in]{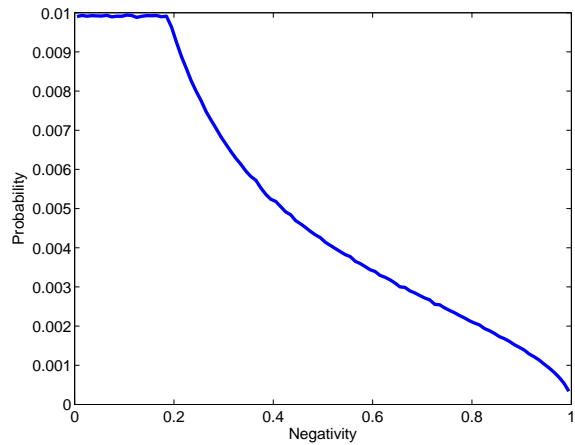}
\caption{Prior probability distribution of the negativity, for the two-parameter states $\rho_{p,\sigma}$ with $p$ and $\sigma$ drawn uniformly from $[0,1]$ and $[0,\pi]$, respectively. The values of the negativity are binned in 100 bins for entangled states. One additional bin is reserved for separable states (of zero negativity). The latter point is not shown in this graph for visual reasons: the probability of separability is 49.7\%. The total number of states drawn from the prior distribution is $N_p\times N_\sigma=600\times600$.
}
\label{negp}
\end{center}
\end{figure}

Let us first discuss the  two-parameter substitute prior with $p$ and $\sigma$ drawn
uniformly from $[0,1]$ and $[0,\pi]$, respectively. As our favorite  entanglement monotone we use the negativity \cite{VidalWerner, NegativityOriginal}.
The prior probability distribution for negativity  is displayed in Figures \ref{negp} (the graph for concurrence is the same for his special case). The plot shows that states exist in the full range of separable to maximally entangled, with the prior probability of entanglement being $P_{{\rm ent}}=50.3\%$.

\begin{figure}[htbp]
\begin{center}
\includegraphics[width=3in]{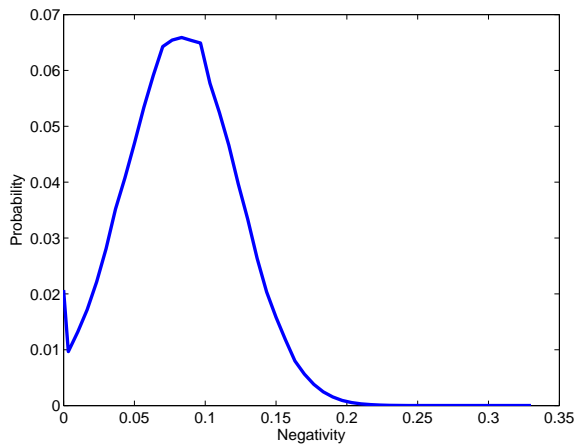}
\caption{The state considered here is of the form $\rho_{p,\sigma}$ with
$p=0.4$ and $\sigma=0.4$. Each of the five Bell correlation observables is measured 400 times, so that in total $N_{{\rm m}}=2000$ measurements have been performed. Plotted is the posterior probability distribution for the negativity. The results have been binned together in 50 bins of equal size for entangled states, plus one extra bin for separable states (at zero negativity). The  posterior probability for entanglement is 98\% in this case (with 2\% falling in the first bin at zero negativity).
The estimated negativity and its error bar are $\bar{N}=0.082\pm 0.039$, where the actual negativity of the state $\rho_{0.4,0.4}$ is $N(\rho_{0.4,0.4})=0.0843$.}
\label{0404}
\end{center}
\end{figure}

\begin{figure}[htbp]
\begin{center}
\includegraphics[width=3in]{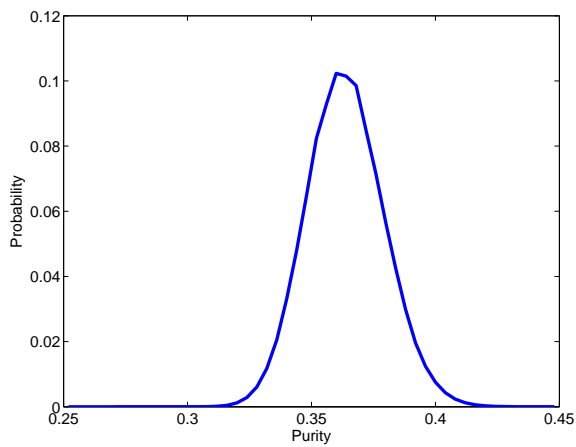}
\caption{Same as Figure \ref{0404}, but for the purity.
The purity of the state $\rho_{0.4,0.4}$ equal 0.3638, the estimate (plus error bar) obtained for the purity is $\bar{P}=0.364 \pm 0.016$. The estimate of the purity is, relatively speaking,  much more accurate than that of entanglement.}
\label{Pur0404}
\end{center}
\end{figure}

\begin{figure}[htbp]
\begin{center}
\includegraphics[width=3in]{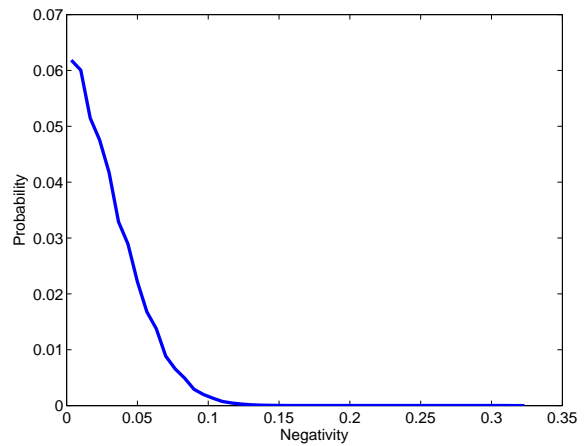}
\caption{Same as Figure \ref{0404}, but for  the separable state $\rho_{1/3,1/3}$. The estimate obtained for the negativity is $\bar{N}=0.012$, and its standard deviation is $\sigma_N=0.020$. The probability of entanglement is $P_{{\rm ent}}=40.5\%$. The bin at zero negativity contains 59.5\% probability, and that point is not plotted for visual reasons.  }
\label{03330333}
\end{center}
\end{figure}

Using this prior, we consider the measurement of five different Bell correlations.
Sample results
are displayed and discussed in Figures~\ref{0404}--\ref{03330333}.
Figure \ref{0404} shows measurement results generated from an entangled state $\rho_?=\rho_{0.4,0.4}$, as defined in Eq.~(\ref{ps1}). There is no need to test either AIC or BIC for this case, since the state is chosen from the two-parameter set of states, so the two-parameter model is trivially more informative.
The Bayesian posterior probability for entanglement distribution is consistent with the actual entanglement properties of $\rho_?$, as discussed in the Figure caption.

We then also test a state that is just separable, the state $\rho_{1/3,1/3}$.
The results can be summarized as ``inconclusive'' about the question whether the data inform the experimentalist that the underlying state is entangled or not. This is not surprising given how close the actual state is to the separable/entangled  boundary. The plot for purity is not shown, as it is very similar to Figure \ref{Pur0404}
(the estimate of the purity is, $\bar{P}=0.331 \pm 0.013$ perfectly consistent with the actual purity of 0.3303 of $\rho_{1/3,1/3}$).

Next we consider the following family of states
\begin{equation}\label{rhok}
\rho_k=0.5|\psi_k\rangle\langle \psi_k|+0.5\openone/4,
\end{equation}
with $k\leq 1$ and
\begin{equation}
|\psi_k\rangle=(|00\rangle+k|11\rangle)/\sqrt{1+k^2}.
\end{equation}
For $k=1$ this state is in the two-parameter set, but for $k<1$ it is not. Obviously, the smaller $k<1$ is, the less well it is approximated by a state $\rho_{p,\sigma}$.
We investigate how well the two-parameter model does by calculating
\begin{eqnarray}
\Delta\Omega&\equiv& \Omega_{p,\sigma}-\Omega_a\nonumber\\
\Delta\Omega'&\equiv& \Omega'_{p,\sigma}-\Omega'_a,
\label{DO}
\end{eqnarray} and tabulating the values for several values of $k<1$ in Table \ref{tab1}.
We moreover give the estimated negativities and purities, plus their error bars, as compared to the actual values of those quantities for the states $\rho_k$.
\begin{table}[htdp]
\begin{center}
\begin{tabular}{|c|c|c|c|c|c|}
\hline
$k$&$N$&$\bar{N}\pm\sigma_N$&$\bar{P}\pm\sigma_P$&$\Delta\Omega$&$\Delta\Omega'$\\\hline
0.9&0.247&$0.246\pm 0.024$&$0.436\pm 0.012$&7.2&36\\
0.8&0.238&$0.237\pm 0.024$&$0.431\pm 0.012$&0.9&30\\
0.7&0.220&$0.219\pm 0.024$&$0.423\pm 0.012$&-11&18\\
0.6&0.191&$0.190\pm 0.024$&$0.410\pm 0.011$&-29&0.8\\
0.5&0.150&$0.149\pm 0.025$&$0.392\pm 0.011$&-53&-24\\
\hline
\end{tabular}
\caption{Comparison, through the Akaike and Bayesian information criteria [using (\ref{DO})], of the two-parameter model based on the family of states $\rho_{p,\sigma}$ [Eq.~(\ref{ps1})] and the
full 15-parameter description of all two-qubit states, with measurement data generated from the family of states $\rho_k$ [Eq.~(\ref{rhok})]. Here the number of measurements is $N_{{\rm m}}=5\times 1000$.
The purity of $\rho_k$ is equal to $P=0.4375$ for any value of $0<k<1$. For decreasing values of $k$, $\Delta\Omega$ and $\Delta \Omega'$ decrease, indicating the two-parameter becomes less and less informative. The estimate of purity becomes, likewise, less and less reliable.}
\label{tab1}
\end{center}
\end{table}
What the table shows is that the two-parameter model ceases to be more informative when $k<1$ decreases.
At that point, the estimate of negativity is still perfectly fine, but the estimate of the purity starts to fail.
In the last entry, for $k=0.5$, the two-parameter's model's estimate of purity is definitely off by a large amount.

\begin{figure}[htbp]
\begin{center}
\includegraphics[width=3in]{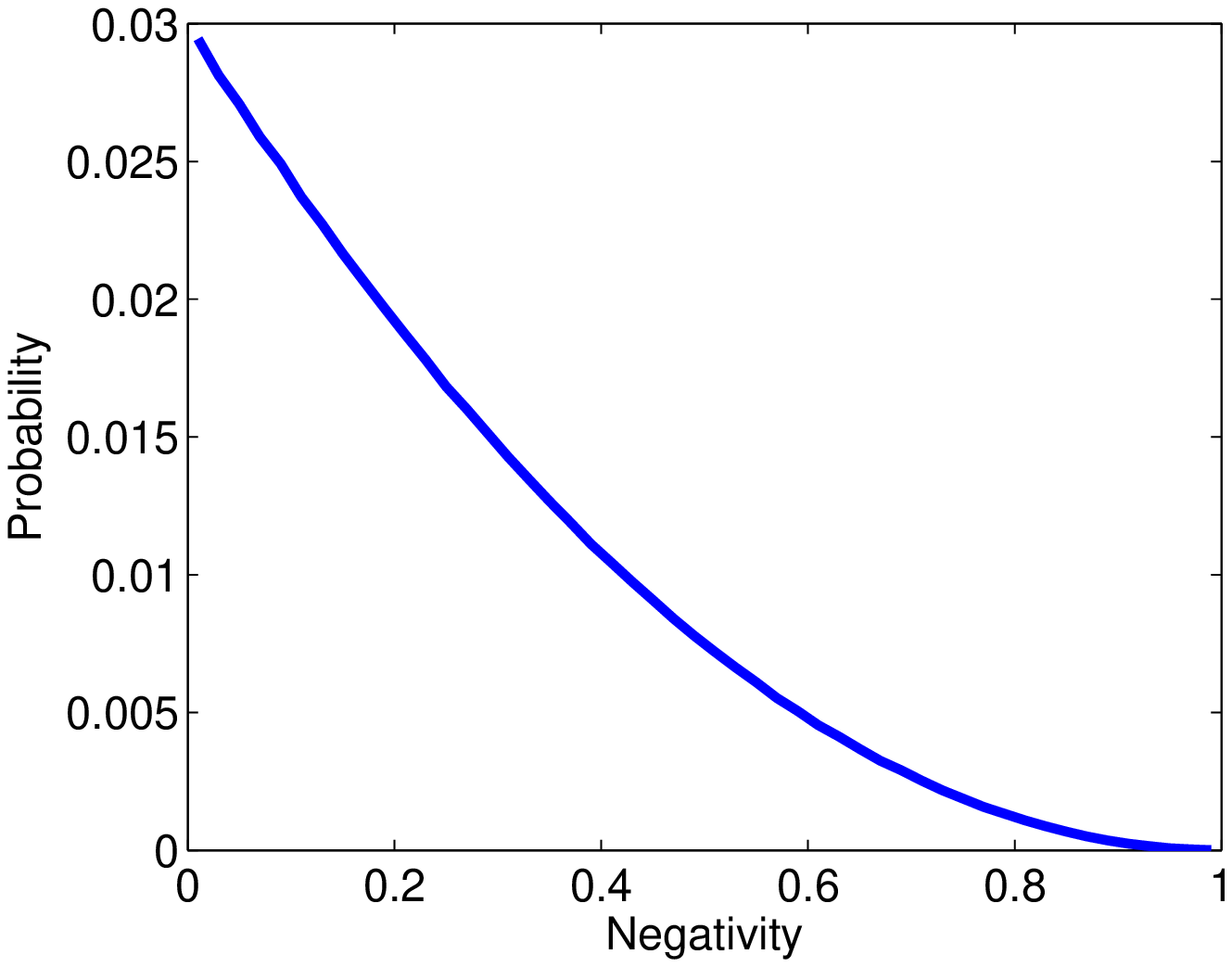}
\caption{Prior probability distribution of the negativity, for the three-parameter set of Bell-diagonal states. The point at zero negativity is left out for visual reason: separable states
occupy 50.0\% of the total volume. Here $10^7$ states were drawn from the prior distribution over states.}
\label{negp3}
\end{center}
\end{figure}

\begin{figure}[htbp]
\begin{center}
\includegraphics[width=3in]{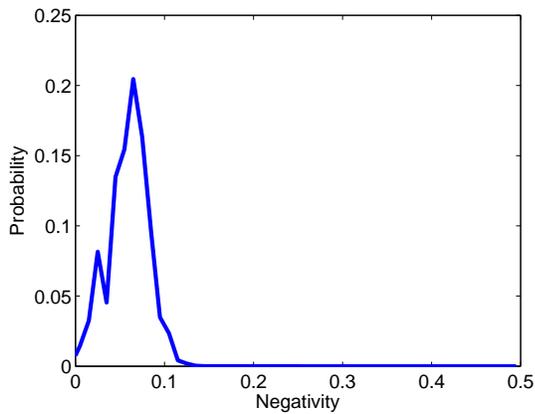}
\caption{Posterior probability distribution of the negativity, using the three-parameter set of Bell-diagonal states as prior, for data generated from the state $\rho_1$ of Eq.~(\ref{rho1}).}
\label{Post3}
\end{center}
\end{figure}

In order to consider the three-parameter Bell-diagonal model, we first display the prior distribution for negativity of that model in Fig.~\ref{negp3}.
Next we us discuss a state that is not close to any state in the two-parameter set of states, but that is still reasonably well described by the three-parameter model,
\begin{equation}\label{rho1}
\rho_1=0.53 |\psi_1\rangle\langle \psi_1|+
0.47 |\phi_1\rangle\langle \phi_1|,
\end{equation}
with
\begin{eqnarray}
|\psi_1\rangle&=&(|00\rangle+0.9|11\rangle)/\sqrt{1.81},\nonumber\\
|\phi_1\rangle&=&(|01\rangle+0.9|10\rangle)/\sqrt{1.81}
\end{eqnarray}
We consider $N_{{\rm m}}=5\times 1000$ measurements, with each correlation being measured 1000 times.
(For the calculations with the three-parameter model, a test set of size $10^7$ was used. In contrast, for the two-parameter model, test sets of size $600\times600$ were sufficient in all cases. This illustrates that choosing a good physical model with as few parameters as possible pays large dividends.)
For this state we calculate the AIC and BIC and compare the two- and three-parameter (Bell-diagonal) models to the full-state model,
\begin{eqnarray}
\Omega_{Bd}-\Omega_a&=&2.4,\nonumber\\
\Delta \Omega=\Omega_{p,\sigma}-\Omega_a&=&-462,
\end{eqnarray}
for the AIC, and
\begin{eqnarray}
\Omega'_{Bd}-\Omega'_a&=&27,\nonumber\\
\Delta \Omega'=\Omega'_{p,\sigma}-\Omega'_a&=&-433,
\end{eqnarray}
for the BIC.
That is, the three-parameter model is considered more informative than the model containing all physical states. On the other hand, the two-parameter model is {\em much} less informative. The estimates for negativity and purity are, for the three-parameter model
\begin{eqnarray}
\bar{N}&\stackrel{Bd}{=}&0.059\pm 0.022\nonumber\\
\bar{P}&\stackrel{Bd}{=}&0.4977\pm 0.0025
\end{eqnarray}
where the actual values are
\begin{eqnarray}
N&=&0.059\nonumber\\
P&=&0.502.
\end{eqnarray}
Thus, both purity and negativity are estimated correctly within the three-parameter model; and this is what one would expect given the AIC and BIC criteria.
The posterior probability distribution for the negativity is plotted in Fig.~\ref{Post3}.

For the two-parameter model, in contrast,
we get
\begin{eqnarray}
\bar{N}&\stackrel{p,\sigma}{=}&0.056\pm 0.025\nonumber\\
\bar{P}&\stackrel{p,\sigma}{=}&0.353\pm 0.009
\end{eqnarray}
so that again the purity estimated by the two-parameter model is way off, although the estimated negativity is still quite good.
Thus, when the AIC and/or BIC criteria tell one not to trust a certain model, it does not imply that {\em all} estimated quantities from that model are, in fact,  incorrect.

Lest one starts to think that the two-parameter model in fact somehow always estimates the negativity correctly, even if the estimated purity is wrong, here is a counter example to that idea: when the $N_{{\rm m}}=5\times 1000$ data are generated by
the  mixture
\begin{equation}
\rho_2=0.53 |\psi_2\rangle\langle \psi_2|+
0.47 |\phi_2\rangle\langle \phi_2|,
\end{equation}
with
\begin{eqnarray}
|\psi_2\rangle&=&(|00\rangle+0.5|11\rangle)/\sqrt{1.25},\nonumber\\
|\phi_2\rangle&=&(|01\rangle+0.5|10\rangle)/\sqrt{1.25}
\end{eqnarray}
whose negativity is $N=0.039$, the two-parameter
model concludes the state is separable with high probability, $P_{{\rm ent}}=3.1\%$,
and $\bar{N}=3.5\times 10^{-4}\pm 0.0025$ (and the estimated purity is incorrect as well: $\bar{P}=0.319\pm 0.008$ instead of the correct value $P=0.502$).
Here, $\Delta\Omega=-203$.

\section{Discussion and Conclusions}\label{conc}

We have demonstrated a method to characterize entanglement sources from finite sets of data, using Bayesian updating for the probability distribution over density matrices. One obtains a {\em posterior} probability distribution for any quantity that can be efficiently calculated from an arbitrary density matrix. For instance, one obtains a probability that one's state is entangled, as well as expectation values of any computable entanglement monotone,  including estimates of statistical errors. These values  should be compared to their {\em a priori} values to judge whether one's measurement results lead one to be more certain about entanglement or less.

For two qubits it is in principle sufficient for the purpose of detecting entanglement to measure spin on each qubit in just two orthogonal directions. On the other hand, empirically, we found that for accurately {\em quantifying} two-qubit entanglement, adding one more correlation measurement is very beneficial. Thus we concentrated on discussing measurements of five spin-spin correlation functions.

It is hard to say in general what sort of measurements, short of fully tomographic measurements, will be sufficient for estimating what sort of quantities.
An easy check, though, is to count by how many parameters a given quantity is determined. For instance, purity is determined by the eigenvalues of the density matrix. Thus for two qubits one needs only three parameters. Thus, reliably estimating the purity of one's output states ought to be easier than estimating entanglement. Our simulations confirm this suspicion, producing relatively smaller error bars for estimates of purity than for entanglement.

It is important to note that in the above we used the phrase ``characterizing entanglement sources,'' rather than ``verifying entanglement,'' because the latter method, in its standard interpretation, has a different meaning: in entanglement verification one tries to find a proof of entanglement convincing a skeptic outsider. But the Bayesian method rather describes one's own belief.
In particular, the difference is that one's prior belief of the entanglement-generating source is certainly to be included in a Bayesian description, but in entanglement verification methods such beliefs are not allowed. Nevertheless, Bayesian methods can be used for the stricter purpose of entanglement verification, as discussed in \cite{Robintbp}.

In order to characterize one's entanglement source, then, it is allowed to use a model describing one's source, based on, e.g., previous experiments and experiences with the same (or similar) device. We provided a criterion to judge whether a given model of one's source is more or less informative than other possible models. In particular, one can always parametrize the output states by using the full
quantum-mechanical description of an arbitrary state of correct Hilbert-space dimension. The latter model,  though, while being complete, may have more parameters than wished for or needed.
Instead, one may be able to use a description of one's source in terms of a (small) number of physically relevant parameters. We proposed to use two criteria to judge the relative merits of such models, the Akaike Information Criterion (AIC), and the Bayesian Information Criterion (BIC) \cite{book}.
We then showed how the AIC and BIC can be used to choose a test set of states i.e., an {\em a priori} probability distribution over quantum states generated by one's source: a Bayesian method, of course, only produces probabilities of entanglement by first choosing a prior.

If a simple model described one's source very well, then one's test set can be based on that model. We applied the AIC and BIC criteria to several examples, all involving two qubits, and showed that indeed, such criteria indicate whether model's predictions about purity and entanglement of the output of the source (including a probability that one's output state is entangled, as well as an estimate of the amount of entanglement) can be expected to be reliable or not. We demonstrated this by showing that certain estimates produced from a simple model are wrong if the information criteria deem the model to be less informative than the full 15d description of two-qubit quantum states, whereas those estimates are right on the mark, when the criteria deem the simple model to be more informative.

\section{Acknowledgements}

SJvE thanks Robin Blume-Kohout for many useful and inspiring discussions. This research is supported by the Disruptive Technologies Office (DTO) of the DNI.

\end{document}